# Unsupervised Data Extraction from Computer-generated Documents with Single Line Formatting

Vladimir Bernstein and Andrei Afanassenkov

vlad@simx.com; andrei@simx.com

SiMX Corporation, July 2020

## Abstract

Processing large amounts of data is an essential problem of the big data era. Most of the data exchange is done via direct communication (using APIs) and well-structured file formats (JSON, XML, EDI, etc.), but a significant portion of the data is transferred using arbitrary formatted computer-generated documents (such as invoices, purchase orders, financial reports, etc.), which require sophisticated processing and human intervention for data interpretation and extraction. The currently available solutions, ranging from manual data entry to low-level scripting and data extraction tools, are costly and require human intervention. This paper describes the principle methodology for unsupervised, fully automatic data extraction from a wide range of computer-generated documents, assuming that their formatting reflects the original structure of the data sources. The presented methodology falls into the category of unsupervised machine learning and consists of the three main parts: (1) - detecting repeating patterns of text formatting by employing the relative feature space clustering and adaptive weighted feature score maps, (2) - detecting hierarchical formatting structures via collapsing and noise filtering procedure applied to the repeating formatting patterns and (3) - automatic configuration of the interactive data extraction tool (SiMX TextConverter) for fully automated processing.

## Introduction

Dramatic growth in information exchange is the hallmark of the 21st century. The most common task in information exchange involves the automated transfer of data from one structured storage to another with minimal or no human involvement. Data flows in a variety of formats, ranging from direct streaming (using standardized protocols and APIs) and well structured file formats (such as JSON, XML, EDI, etc.) to a variety of documents with arbitrary layouts.

The first two types of data exchange (direct streaming and predefined formats) are easy to automate and require limited human interaction, while the last one (via arbitrarily formatted documents) needs either extensive human intervention or sophisticated technology in order to interpret and structure the document's content. At the same time, a big part of such documents are programmatically produced (i.e. computer-generated) and therefore have consistent internal structure, which provides an opportunity to automatically recognize this structure and use it for data interpretation and extraction. In this paper we will consider the theoretical and practical sides of the methodology for solving this problem.

Before getting to the main subject, let us consider the available solutions, which could be divided into four categories:

1. Manual data entry
2. Low level scripting
3. Supervised document processing using data-extraction software tools
4. Supervised document processing using AI technology
5. Unsupervised document processing using AI technology

Manual data entry (1) is predictably time-consuming and error-prone (but still widely used due to simplicity of implementation). Scripting (2) offers reliability and efficiency, but is expensive to implement and manage, requiring coding skills and deep understanding of the data. The supervised document processing using data-extraction tools (3) (such as Altair Monarch, Astera ReportMiner, docparser or SIMX TextConverter) is less expensive, but still requires extensive human intervention.

In spite of its rapidly growing capabilities, most of the practical achievements of AI technology are concentrated on supervised machine learning (4). For example, the most known AI-based document processing solutions are offered by Google and Microsoft. The detailed analysis of such solutions goes beyond the scope of this paper, but we will briefly review one of the most advanced examples offered by Google: the Entity Extraction component of Google's AutoML Natural Language cloud services.

While Google's deep learning methodology may potentially work autonomously in the future, the currently available version requires extensive human participation, primarily to configure the training phase. During this phase, Google recommends adding at least 200 annotations for each extracting field, which for the sample input document (considered in this paper - financial report with about 30 fields), led to several hours of intensive interactive work. With the addition of the AI model's training time, configuring this type of job takes about 10 hours per format.

These results make the Google AutoML Natural Language data extraction solution more efficient than manual data entry, comparable to low-level scripting, but still inferior compared to data extraction software tools, which typically take a few hours to configure and consume dramatically less computing power. The main subject of this article falls into the category of unsupervised document processing employing artificial intelligence and machine learning technology (5) that seeks to address these deficiencies.

## Task Definition

The process of generating documents programmatically makes it reasonable to assume that, unless some specific steps have been taken to break the order, the formatting of computer-generated documents reflects the structure of the source data. For example, the typical structure of the invoices produced by the same process would consist of two parts: invoice related data (invoice number, payor and provider information, etc.) and items data (item description, number of units, unit and item prices, etc.). Each part would be generated by processing records from the corresponding database table (or record stream) where each record would be processed by the same document producing algorithm, generating the same formatting.

The introduced assumption of direct correlation between the computer-generated document formatting and the source data structure opens the way to automatic data extraction, which would consist of two parts:

1. Automatic recognition of the document's formatting structure
2. Automatic configuration of the data extraction project.

The first task comes to detecting identically formatted text elements, identifying the repeating patterns of their occurrence in the document flow and reconstructing the hierarchical logical structure(s) linking them to each other. This task falls into the category of pattern recognition and specifically - unsupervised machine learning, which assumes relying exclusively on the internal information contained in the input data flow.

The second part could be implemented using a low level programming or automated configuration of some interactive software tool via API. The last approach provides a number of advantages (including higher manageability and lower cost), therefore, as an example, we described the process of configuring the data extraction tool [SIMX TextConverter 4](#).

Although this paper is mostly dedicated to a specific type of "single line formatting" documents, let us first consider a simplified, but more generic example of a three-page document containing two invoices with free layout and "multiline formatting". This should help to better understand the basic concept of the considered task and the scope of the offered methodology applicability.

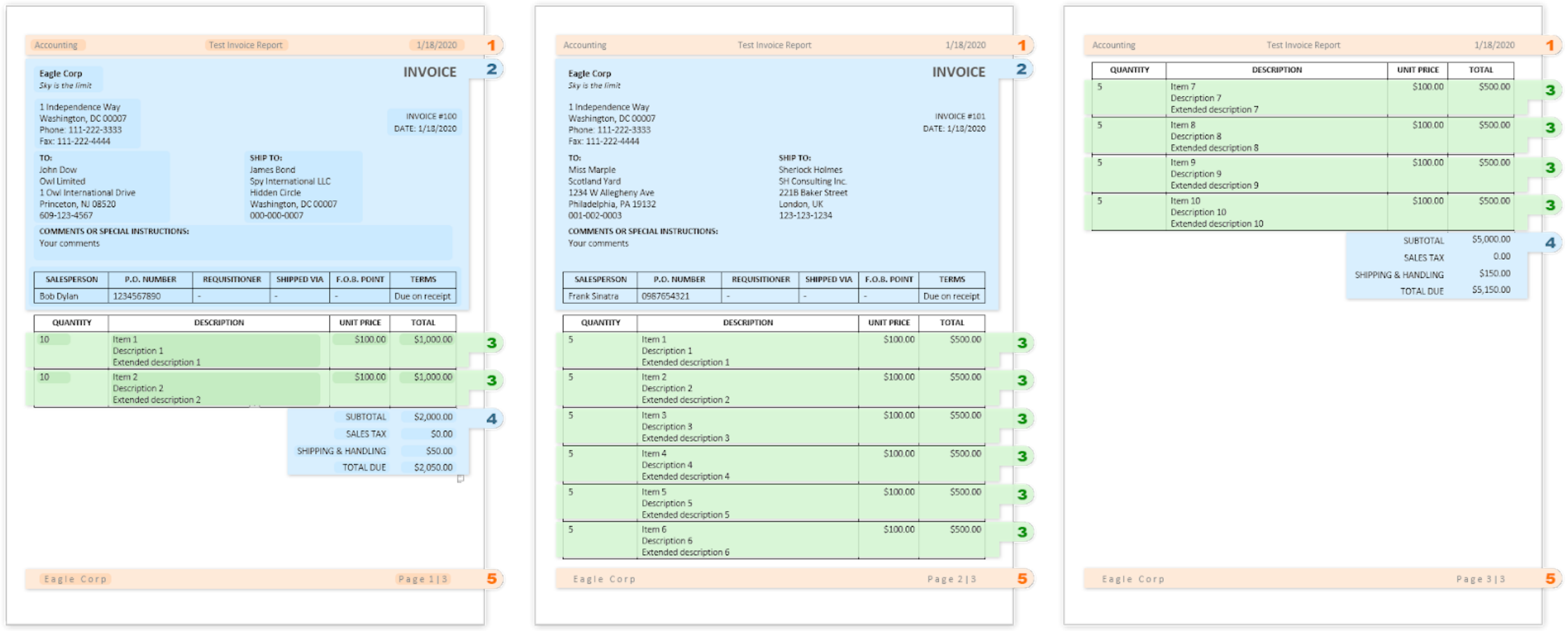

Figure 1. Sample invoice document with the color highlighted repeating formatting sections (patterns).

The colored highlighting of the document sections indicates the layout/formatting structure of the document that we are looking to discover. The pink sections (1 and 5) mark the page header and footer areas; the blue colored areas (2 and 4) represent the invoice header and footer sections and the green sections (3) contain the detailed items related data.

By assigning sequential identification numbers to the uniquely formatted repeating sections (formatting patterns) and putting them together in the order of their appearance in the document flow, we can turn the description of the document's formatting structure into a simple number series:

1,2,3,3,4,5,1,2,3,3,3,3,3,3,5,1,3,3,3,3,4,5

where each number corresponds to a unique repeating pattern of formatting. The Figure 2 below shows the produced number series as a chart (the background color indicates the sections' role in the document).

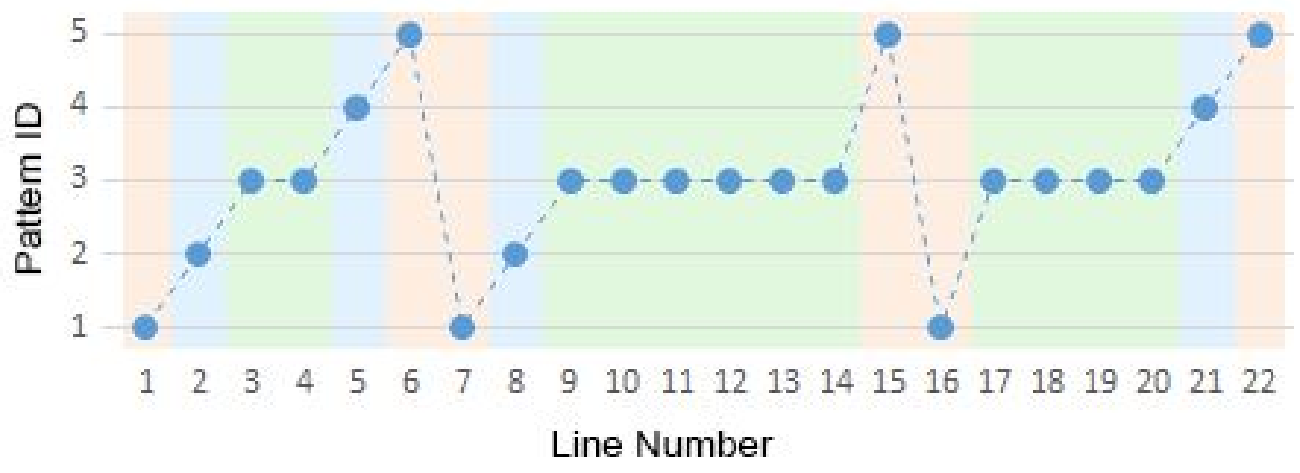

Figure 2. Visualization of the sample invoice document formatting patterns' number series.

The logic of the report generating process allows us to interpret the content of this number series as two independent overlapping structures, described in the table below.

| Structure | Level | Pattern | Role |
|---|---|---|---|
| 1 | 0 | [3] | Invoice items (details) |
| | 1 | [2, [3], 4] | Invoice header and footer |
| 2 | 0 | [1, 5] | Page header and footer |

Table 1. Sample invoice document formatting structure description.

The first structure (Structure 1) consists of two nested hierarchical levels of patterns (Level 0 and Level 1) representing the invoice related data (Pattern [3] and Pattern [2, [3], 4] and the second structure (Structure 2) represents the page headers and footers (Pattern [1, 5]). Each pattern consists of a combination of templates (including the repeating pattern of the previous level). We can also present the entire document structure as a single string: "[2, [3], 4] / [1, 5]", where numbers represent formatting templates, brackets indicate inclusion of the hierarchical levels, and the 'slash' - separates different hierarchical structures from each other. Automatic generation of such a string (let us call it 'Document Structure String' or DSS) for any arbitrary computer-generated document would be the first goal of our analysis.

Each repeating pattern in this example consists of a number of specifically formatted text segments representing the source database fields of the same record, which belongs to a certain level of one of the hierarchical document data structures. It leads to the following rule enabling detection of repeating patterns of objects in the document's text flow: *objects with similar formatting should belong to the same level of the hierarchical structure formed by those objects*.

In other words, the overall task of automatic recognition of the document data structure consists of the three interdependent subtasks:

1. Detecting text objects with similar formatting.
2. Detecting repeating patterns of such objects occurring in the text flow.
3. Detecting the hierarchical structure formed by the detected repeating patterns.

Although this methodology is applicable for documents with arbitrary layout, in this paper we will consider the limited range of documents with one line formatting patterns (which represents a compelling portion of the document types), thus significantly reducing complexity of the first subtask.

As a working example, we will use the financial report below, which has a more complicated internal structure (compared to the invoices sample above), but simplified single line formatting patterns. Each repeating pattern could consist of multiple one-line formatting templates. Let us start with defining the expected outcome of the automated analysis based on our preliminary expert-based knowledge of the document's structure (as we did with the invoice example above).

The document consists of the three-level hierarchical structure representing the actual reporting data and the one-level page header and footer.

```
Line                                                                                                                           Template
 1   REPORT: MCDRA03                         INVOICE VARIANCE REPORT                                          PAGE:   1          0
 2  PROGRAM: ED89510                          WALGREENVA  INV SW                       DATE: 09-20-17     TIME: 11:46:33         1
 3                                                    UPC/           INV                  CUSTOMER  MCLANE                       2
 4  DIV CUST-NUM STORE  INV-VAR-ID          PO-NUM    DESC           QTY   UM PACK  PRICE   VALUE     VALUE      VARIANCE   UM ADJ TYP  3
 5  --- -------- ------ ------------------- --------- -------------- ----- -- ----- ------- --------- ---------  ---------  -- -------  4
 6  SW  01667949 004513 0130687732I         337261                                                                                   5
 7                                                    02610080575        8 CT    10   63.04    2.0000    8.0000    -6.0000  CT SHORT    6
 8                                                    292235 NEWPORT MTL BX   FSC                                                    7
 9                                                    02720001865        4 CT    10   54.74    2.0000    4.0000    -2.0000  CT SHORT    6
10                                                    358416 PALL MALL MTHL 100 BX FSC                                               7
11                                                    02820019830        1 CT    10   62.44              1.0000    -1.0000  CT SHORT    6
12                                                    738708 VIRG SL SS GOLD MTL BX FS                                               7
13                        TOTAL INVOICE ADJUSTMENT    ( -550.16)                                                                     8
14  SW  01667949 004513 0130687734I         337259                                                                                   5
15                                                    04902264490       15 EA    15   17.68             15.0000   -15.0000  EA SHORT    6
16                                                    065136 NICE LRG GRADE A EGGS                                                   7

49                                                    04902295576        2 EA     1    1.13              1.1300     0.1200  EA PRICE    6
50                                                    936484 NICE RST N SLT MX NUTS SS                                               7
51                        TOTAL INVOICE ADJUSTMENT    (  -16.76)                                                                     8
52  SW  01670224 007671 0130712795I         219546                                                                                   5
53                                                    02590020748       30 EA    15   10.96              0.7306     0.0004  EA PRICE    6
54                                                    542100 WLGRNS SS GRAPE CGRLO PCH                                               7
55   REPORT: MCDRA03                         INVOICE VARIANCE REPORT                                          PAGE:   2          0
56  PROGRAM: ED89510                          WALGREENVA  INV SW                       DATE: 09-20-17     TIME: 11:46:33         1
57                                                    UPC/           INV                  CUSTOMER  MCLANE                       2
58  DIV CUST-NUM STORE  INV-VAR-ID          PO-NUM    DESC           QTY   UM PACK  PRICE   VALUE     VALUE      VARIANCE   UM ADJ TYP  3
59  --- -------- ------ ------------------- --------- -------------- ----- -- ----- ------- --------- ---------  ---------  -- -------  4
60                        TOTAL INVOICE ADJUSTMENT    (   -0.01)                                                                     8
61  SW  01670943 009697 0130715273I         215694                                                                                   5
62                                                    02590020748       30 EA    15   10.96              0.7306     0.0004  EA PRICE    6
63                                                    542100 WLGRNS SS GRAPE CGRLO PCH                                               7
64                        TOTAL INVOICE ADJUSTMENT    (   -0.01)                                                                     8
65                      DIVISION TOTAL ADJUSTMENTS    (  -891.68)                                                                    9
```

Figure 3. Sample computer-generated text document with the highlighted hierarchical sections.

Where:

| | |
|---|---|
| green | - the lowest (detailed) level, |
| light blue | - the first summary level contains the group header and footer, |
| dark blue | - the second summary level contains only a footer, |
| pink | - page header and footer. |

On Figure 3, the numbers to the left indicate the line numbers and to the right - the templates' identification numbers.

The number series on the Figure 4 shows the unique formatting patterns as a function of the order of their appearance in the document flow (which due to the one-line formatting limitation matches the line numbers).

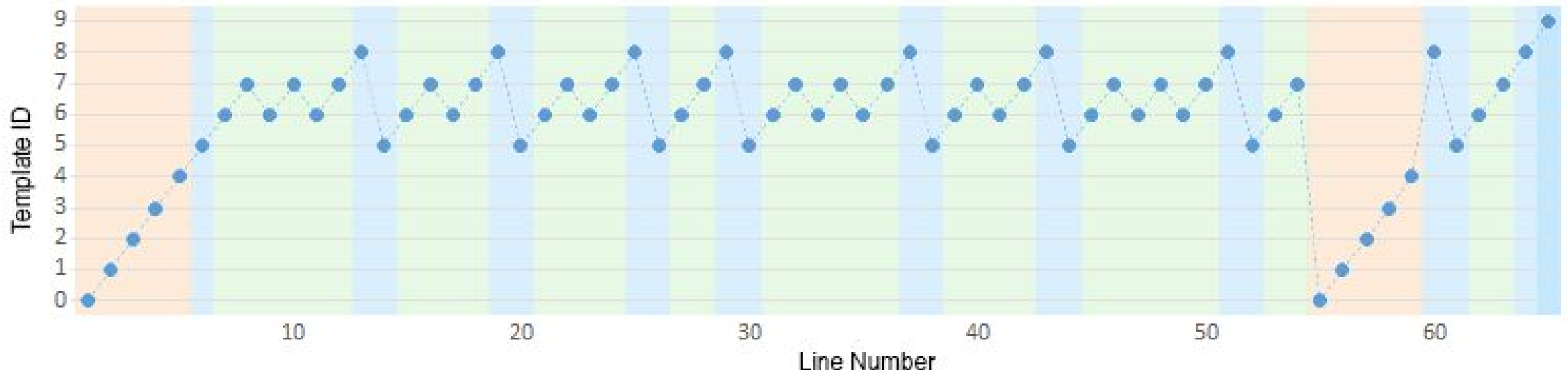

Figure 4. The number series of the unique formatting patterns shown as a function of line numbers.

The desired results of our automated analysis are presented in the table below. They include the two structures, first of which consists of the three nested hierarchical levels of repeating patterns and the second contains only one. Each pattern consists of a combination of templates, which include the patterns from the previous levels. The Document Structure String (DSS) representation of this document's structure would be: "[[5, [6, 7], 8], 9] / [0, 1, 2, 3, 4]"

| Structure | Level | Pattern | Role |
|---|---|---|---|
| 1 | 0 | [6, 7] | Details level |
| | 1 | [5, [6, 7], 8] | Summary level 1: 5 - group header, 8 - group footer |
| | 2 | [[5, [6, 7], 8], 9] | Summary level 2: 9 - group footer |
| 2 | 0 | [0, 1, 2, 3, 4] | Page header |

Table 2. Sample financial report document formatting structure description.

For most types of documents the input data flow could be presented as two types of streams:

1. Characters (including line separators)
2. Groups of characters equipped with their relative or absolute positioning coordinates.

In this paper we will focus on the first type of data streams, which is relevant for all types of text files and a majority of the popular document formats that could be converted into a text stream. Also, as it was mentioned earlier, at this point we will not consider documents with multiline fields. Thanks to the made assumptions, the input data flow can be presented as a stream of text lines.

As it was mentioned earlier, the task of recognising computer-generated documents' structure includes the three main interdependent components:

1. Detecting identically formatted objects (text lines in our simplified sample).
2. Identifying repeating patterns of objects' occurrence in the document flow.

3. Reconstructing the logical structure of the document.

As it will be shown later, these components depend on each other, which makes it necessary to eventually employ some iterative or optimization process that would integrate them into a single process. Keeping this in mind, we will consider each of the components separately, before merging into a final integrated solution.

## Detecting identically formatted text lines

In this section we will consider the task of detecting lines with similar formatting in the flow of input text lines. This task falls into the category of unsupervised machine learning and more specifically - unsupervised objects' classification. The most relevant conventional methodologies for solving this problem would be the unsupervised clustering methods [1], such as K-Means [2], Hierarchical clustering [3] or DBSCAN [4].

Most of the clustering methods are based on defining a basic object type by selecting a number of observation measurements (features) that would allow identifying those objects in the input data flow, presenting them as points (or vectors) in the multidimensional feature space and using various algorithms for grouping points into classification clusters. The defined features are assumed to be static, which makes the feature space universal for the entire input data flow.

In our case, the basic object type of interest would be the text line with specific formatting identifiable by a properly selected set of features. We will also keep in mind that the identified line objects should form a multilevel hierarchical structure that will be defined later.

### *Formatting features*

Selecting the proper feature set (identifying the basic object type) is crucial for the recognition outcome, but due to the lack of objective criteria, it is usually selected intuitively, with possible corrections based on the subjective estimation of the achieved classification results.

In our case of identifying lines with identical (or similar) formatting, the central formatting feature would be the repeating (matching) positioning of the blocks of text presumably associated with the source database fields. These blocks (let us call them fields) could consist of one or more words, which borders' position, depending on the alignment, could represent the left or right border of the field. Repeating the same field border position in different lines would be one of the most distinguishing line formatting features. This feature could be detected as a combination of a space and non-space characters: the leftmost space (or beginning of line) - for the left border and the rightmost space (or end of line) - for the right border. Another important feature would be the positioning of the specific characters and character types inside the words or on their borders.

Using the uniform feature space assumes that the input objects (text lines in our case) could be uniquely identifiable by a set of static features (such as the ones described above). This makes it necessary to associate the introduced features with each character position of the line, leading to an excessive number of feature space dimensions equal to the number of characters in the longest line. For example, the unique identification of the two character lines with the formatting features defined as a character type (1 - alpha and 2 - numeric), would look like this:

| Line # | Text | ID Vector |
|---|---|---|
| 1 | **AB** | {1,1} |
| 2 | **C1** | {1,2} |

Table 3. Sample unique identification of the two character lines.

The conventional feature space approach assumes that the distance between points (Euclidean or other) represents the dissimilarity of the corresponding objects. The feature space for the two character lines described above would look like this:

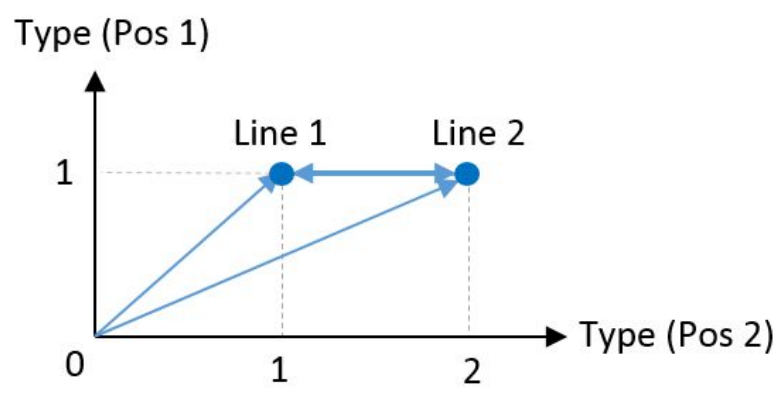

Figure 5. Universal dissimilarity feature space for the two character lines.

As it was shown above, using the universal feature space for comparing text lines would require an excessive number of dimensions (133 for our sample document), thus increasing the complexity of the task and the required computing power. At the same time, our task of identifying the logical structure of lines contained in one document is limited to the relatively small number of line formatting types and therefore, does not require a universal (absolute) identification of lines.

Instead of calculating the universal *dissimilarity* distance between the points, the task could be solved by calculating the distance between the selected and other points in the local (relative) *similarity* feature space attached to the selected point. Using such a relative feature space allows to dramatically minimize the number of considered features to only those relevant to the specific context and to replace the Euclidean distance calculation with a simple sum of vector lengths, thus reducing the number of feature space dimensions to only one. This approach makes it easier to implement the flexible algorithm calculating the lines' *similarity score* based on the adaptive weighted score maps.

Using relative feature space allows to implement the lines classification procedure by sequentially comparing lines, one by one to all the others, employing the specific features relevant to each comparison context. This approach also resolves the problem of determining the number of classification clusters. Each iteration of comparing the selected line to all the others splits the lines into two clusters: the ones with the formatting similar to the selected line and all the others. The similarly formatted lines would be marked as recognized and associated with a specific format id (template). The process continues until there are no unmarked lines left, thus resulting in the final number of clusters.

As it was mentioned above, the identically formatted lines could be classified differently depending on their position in the document's logical structure. For example, any two identically formatted lines located in the *detailed* section of the document hierarchy could be considered as a part of the same class, while otherwise they should be recognized as different ones.

### *Feature Score Map*

In our case, the relevant formatting features would include the absolute or relative positioning of specific combinations of characters or character groups, such as beginning and ending of text blocks or matching sequences of characters or character types (alpha, numeric, symbols), etc. To further simplify the task we assume that text does not scale or shift within the same document (reasonable for most of the computer-generated documents).

The key indicator defining the lines format similarity is the close positioning of the same formatting features in the compared lines. The assumption regarding the lack of text shifting (introduced above) allows to replace the more complicated calculations of the relative difference in positioning of the same formatting features with the exact matching.

Depending on the task, the set of relevant features and their influence on the outcome might vary, therefore we developed a *comparison score calculating procedure* utilizing a flexible weighted feature map (Feature Score Map) that could be adjusted for different tasks and contexts. The Score Map, as we define it, is a matrix of rewarding scores for 15 different format-related events that could be detected while comparing two lines of text. The default Feature Score Map below was produced empirically after processing several hundred documents.

| Character Group | Body Full Match | Body Group Match | Border Full Match | Border Group Match | Border no match |
|---|---|---|---|---|---|
| Alpha | 3 | 3 | 5 | 5 | 1 |
| Numeric | 3 | 5 | 7 | 5 | 1 |
| Symbol | 8 | 6 | 9 | 5 | 1 |

Table 4. Sample Feature Score Map.

In this table: Body – an internal part of a word (not including first and last characters); Match – an identical (or close) absolute position in the line; Border – the first or the last character of a word; Alpha – an alphabetic type of a character; Numeric – numeric type of a character; Symbol – symbolic type of a character.

The content of the Feature Score Map defines the sensitivity of comparison results to specific formatting features, thus providing an ability to fine-tune the procedure towards the specific contexts. The implemented score-calculating algorithm gives normalized results that depend on the relative difference between the elements of the Feature Score Map and not on their absolute values or length of the compared lines.

The image below shows how the variation (from 1 to 9) of each of the 15 Feature Score Map elements affects the total comparison score for the two similar lines from the detailed section of the sample report document above (lines 7 and 9).

```
STR1:                                                               02610080575        8 CT    10   63.04     2.0000    8.0000     -6.0000  CT SHORT
STR2:                                                               02720001865        4 CT    10   54.74     2.0000    4.0000     -2.0000  CT SHORT
```

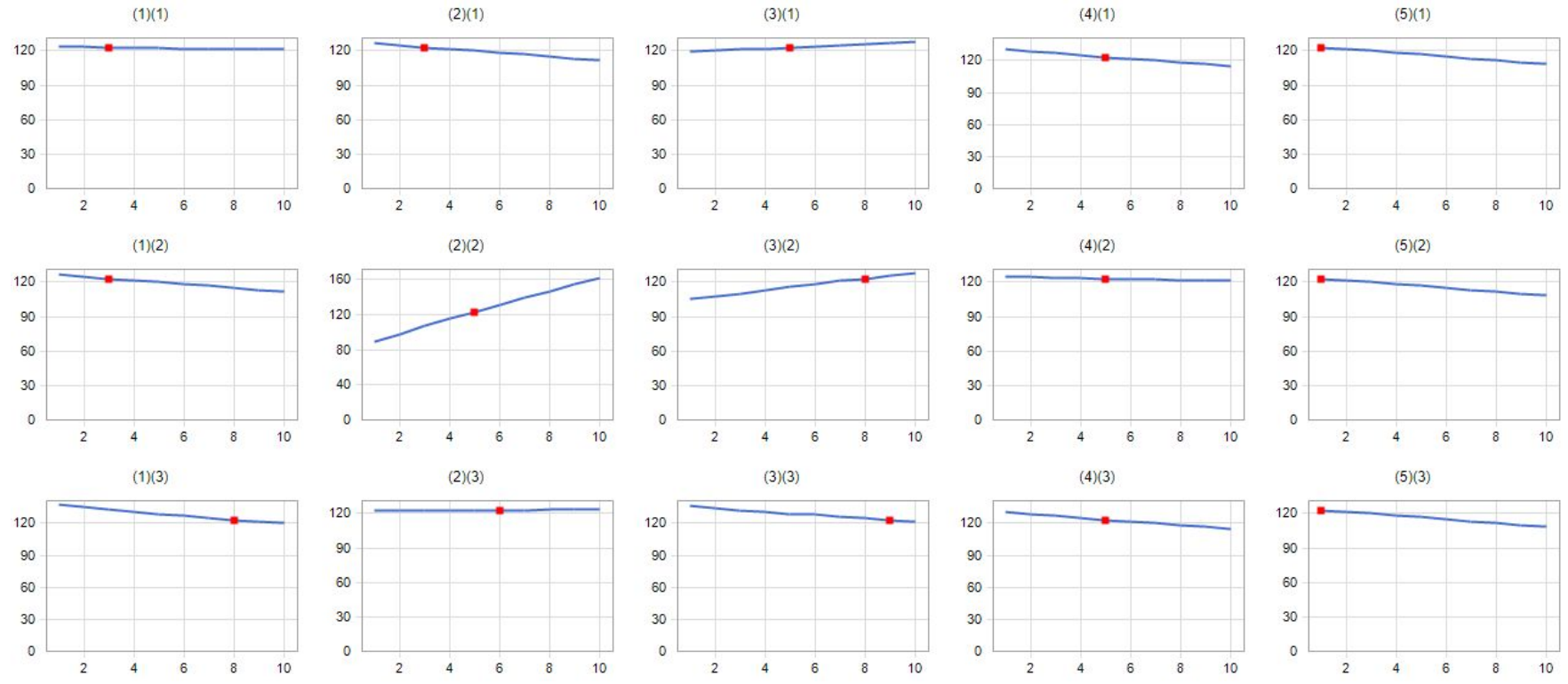

Figure 6. Default Feature Score Map variation results for two text lines with similar formatting.

The calculated total score for the default Feature Score Map equals to 123 and the variation range comes to 75 - 170. The chart shows that the most influential Feature Score Map element is (2)(2) – "Body Group Match for Numerics" (matching positioning of the numeric type characters located inside the words). This happens because the compared lines mostly consist of numeric characters.

The results of comparing lines 7 and 8 (image below), which are not to be recognized as "similar", gives us the total score of 18 for the default Feature Score Map and the variation range from 7 to 30.

```
STR1:                                     02610080575        8 CT    10   63.04    2.0000    8.0000    -6.0000  CT SHORT
STR2:                                     292235 NEWPORT MTL BX    FSC
```

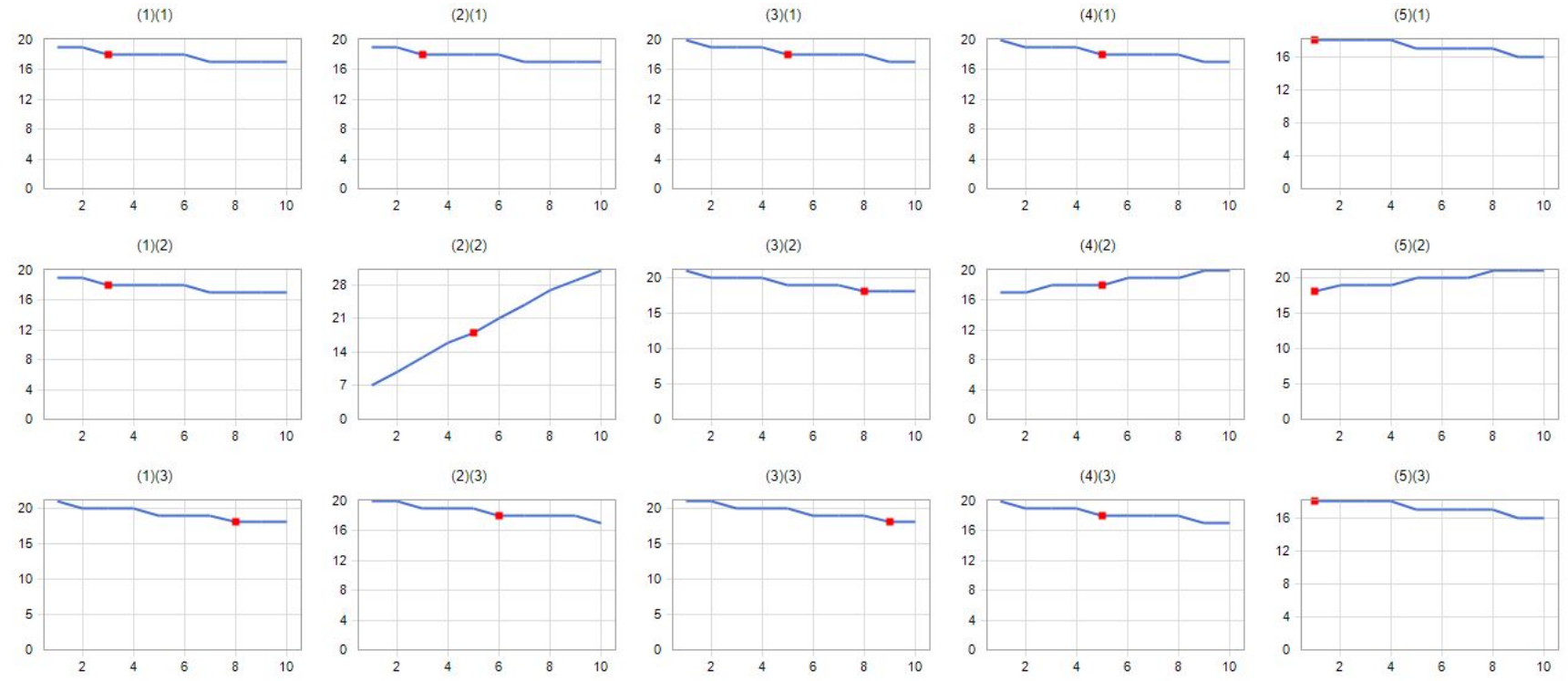

Figure 7. Default Feature Score Map variation results for the two lines with different formatting.

Using the default Feature Score Map with variable elements one at a time produced promising results. Now, let us see how sensitive the lines comparison algorithm will be to changing the entire Feature Score Map.

Applying the “even” Feature Score Map with all its elements set to 5 (as an example) for the first case (comparing lines 7 and 9 of the sample document) gives the total score of 100 with the variation range from 60 to 130. For the second case (comparing lines 7 and 8) it gives the total score of 20 with the range from 6 to 32.

This outcome demonstrates that the suggested method of estimating the lines formatting similarity for the selected example provides reliable and consistent results for a wide range of the Feature Score Map parameters. The not overlapping ranges of the total score for “similar” and “different” pairs of compared lines provides a reliable difference between the two (over 80% for the default Feature Score Map).

The next logical step in our research would be to apply the suggested score-calculating algorithm to the sample document for finding the lines with similar formatting that supposedly were produced by processing the same database record stream.

### *Recognising Templates*

Let us start with comparing the first line with the following 99 lines of the same document. The left chart below shows the results of this comparison as a function of the line numbers and the right chart represents the sorted list of the found unique scores.

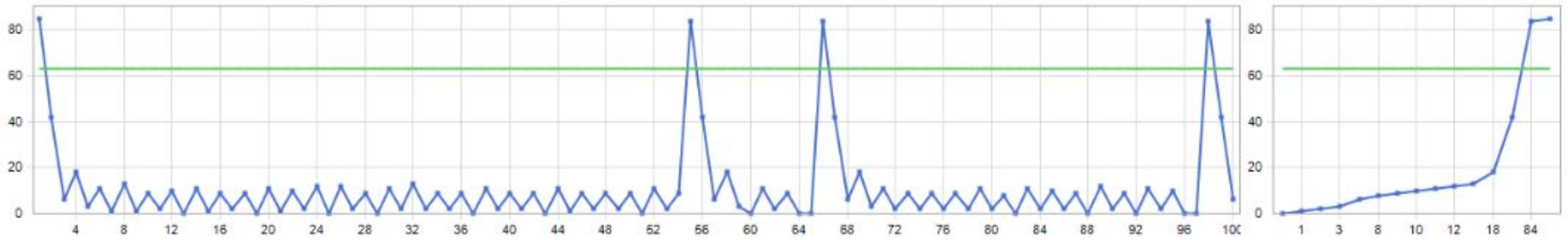

Figure 8. Score calculating results for the first 100 lines compared with the first one using the default Feature Score Map.

The four outstanding score spikes (lines 1, 55, 66 and 98) represent the first lines of the page header. Identifying these lines as a part of the group associated with the unique formatting pattern (“Template 1”), would be the correct outcome for this step. It could be achieved using a *recognition score threshold* with the value in between 44 and 87 that would separate those 4 points from the rest.

The gap between those score points is 43 and it is the biggest gap in the number series, which would be the logical base for separating the points with the closest to each other's highest scores. We used the middle of this gap for setting the recognition threshold (green line).

The size of the biggest gap appears to be about 40% higher than the next biggest gap of 25 points (between 19 and 44, where the score of 44 points represents the second line of the page header). Although, in this particular case the suggested methodology of defining the recognition threshold led to the correct result, the 40% reserve does not look like a reliable one.

As it was mentioned above, the suggested format recognition method could be configured to be more sensitive to specific formatting features by modifying the corresponding Feature Score Map matrix. So far, we used the default version of the Feature Score Map, configured for working with a wide range of generic formatting features. At the same time, the first application of the default Feature Score Map revealed a number of lines with the similar formatting, which allows to identify the specific features relevant for this particular context and use this information for fine-tuning of the Feature Score Map.

### *Adaptive Feature Score Map*

Let us look at the default Feature Score Map variation results (below) for comparing the two most similar lines of the recognized group: lines 1 and 55.

```
STR1:   REPORT: MCDRA03                           INVOICE VARIANCE REPORT                                   PAGE:   1
STR2:   REPORT: MCDRA03                           INVOICE VARIANCE REPORT                                   PAGE:   2
```

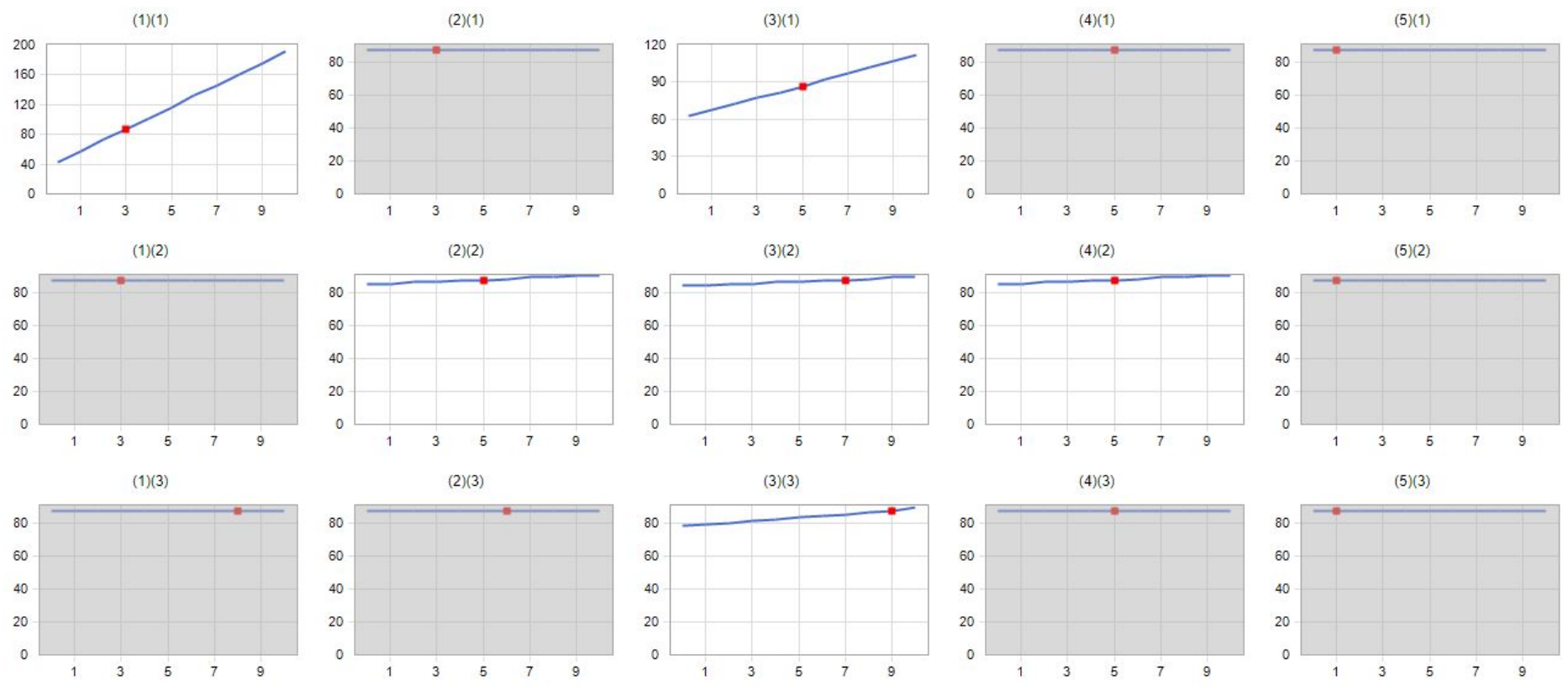

Figure 9. Score variation results for comparing the first line with the most similar one using the default score map.

Each of the charts above represents the variation of the total comparison score as a response to a particular element of the Feature Score Map variation. A closer look at the charts reveals that not all of the Feature Score Map features can influence the total score. Changing the grayed out items (1)(2), (1)(3) , (2)(1), (2)(3) , (4)(1), (4)(3), (5)(1), (5)(2) and (5)(3) did not affect the output score. The obvious conclusion is that those Feature Score Map elements could be set to zero thus increasing the relative weight of the remaining ones.

$$\begin{pmatrix} 3 & 3 & 5 & 5 & 1 \\ 3 & 5 & 7 & 5 & 1 \\ 8 & 6 & 9 & 5 & 1 \end{pmatrix} \Rightarrow \begin{pmatrix} 3 & 0 & 5 & 0 & 0 \\ 0 & 5 & 7 & 5 & 0 \\ 0 & 0 & 9 & 0 & 0 \end{pmatrix}$$

Below are the results of comparing the same two lines using the modified (adapted) Feature Score Map.

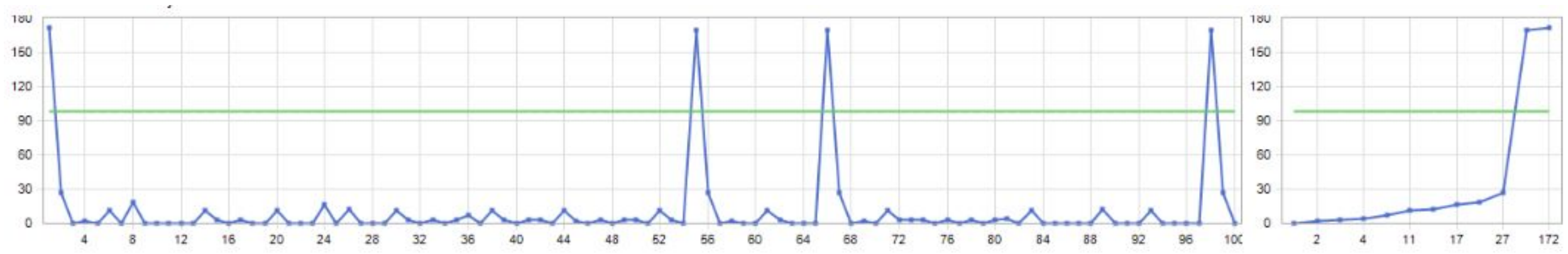

Figure 10. Score calculating results for the first 100 lines compared with the first one using the adapted Feature Score Map.

This outcome provides a much higher difference between the biggest and the next in size gaps between the score points (94% vs 43%) thus significantly increasing reliability of the recognition process.

### *Detecting Templates*

The next step would be to apply the described format recognition mechanism to a representative part of the document in order to identify and mark the lines with "similar" formatting, generate the corresponding templates and collect the data fields-related information that could be used later for the data extraction.

For that, we will compare each of the first 200 lines of the sample document (going from top down) with the succeeding lines (skipping already marked), each time creating a new template and marking the recognized lines with the template number, thus excluding them from further processing. We will continue until all the lines of the selected sample part are marked. In the process we will collect the statistical information that later could be used for configuring automated data extraction projects. The next chart demonstrates the results of such processing.

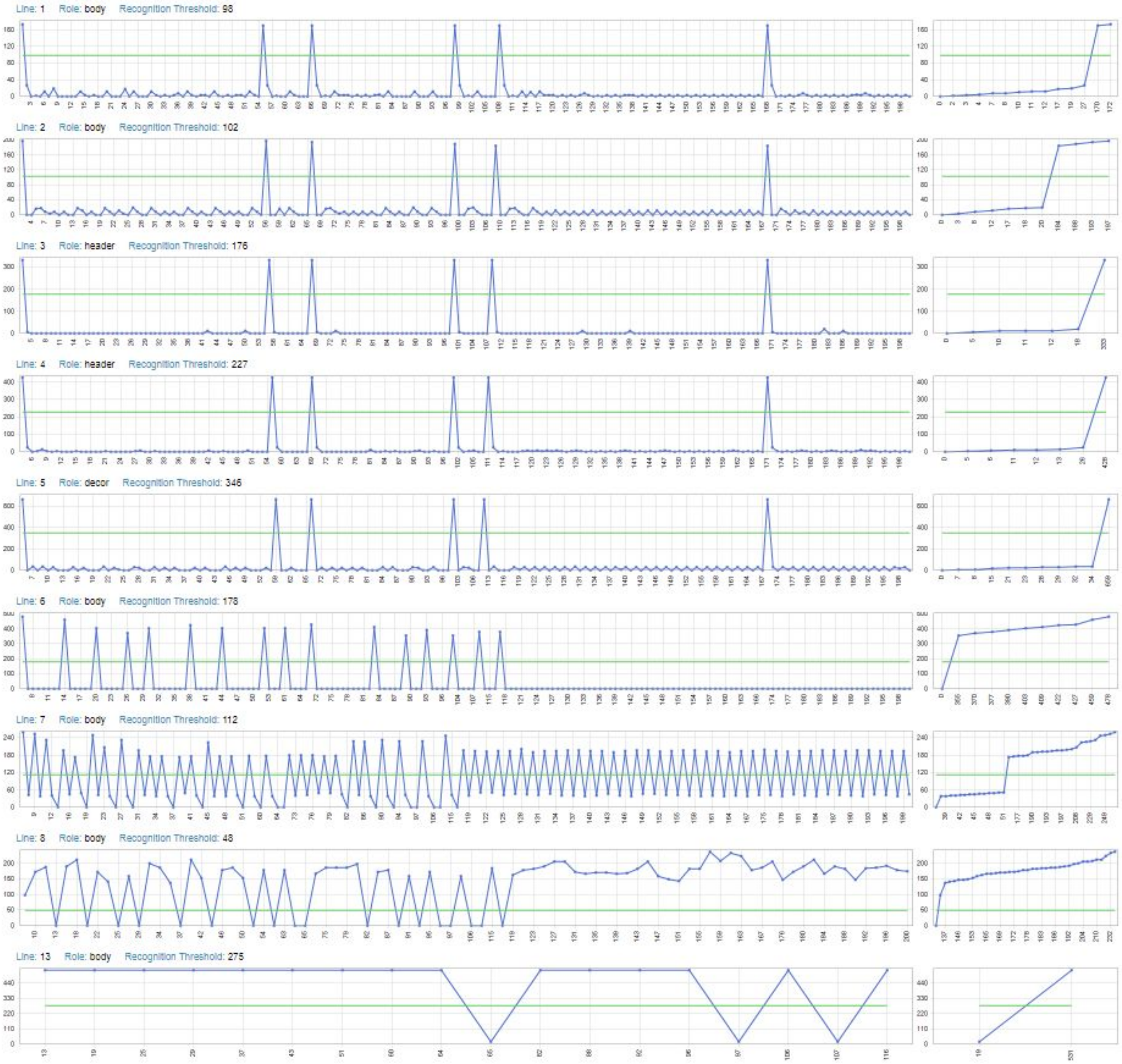

Figure 11. Score calculating results for the complete template recognition process.

The process produced 10 templates (id: 0 – 9) with the recognition threshold ranging from 48 to 346. The templates recognition process also collected the additional information needed for configuring the data extraction projects, such as statistical description of the line mask (regular expression for each character position).

Below is the list of generated templates with the template id, key name (created out of the field positions), template role (detected by analyzing the line's content), number of fields and number of lines associated with this template.

| Template id | Template Key Name | Role | Fields Number | Line Count |
|---|---|---|---|---|
| 0 | 2-10-44-52-61-120-128 | body | 7 | 6 |
| 1 | 1-10-45-57-61-96-102-115-121 | body | 9 | 6 |
| 2 | 53-68-91-101 | heading | 4 | 6 |
| 3 | 1-5-14-21-42-53-68-74-77-83-92-101-112-123-126-130 | heading | 16 | 6 |
| 4 | 1-5-14-21-42-53-68-74-77-83-91-101-112-123-126 | decor | 15 | 6 |
| 5 | 1-5-14-21-42 | body | 5 | 16 |
| 6 | 53-70-74-80-85-94-102-112-123-126 | detail | 10 | 68 |
| 7 | 53-60 | detail | 2 | 68 |
| 8 | 24-30-38-53-55 | body | 5 | 15 |
| 9 | 22-31-37-53-56 | body | 5 | 3 |

Table 5. Recognized line Templates

## Detecting hierarchical structure of repeating patterns

### *Hierarchy in the templates number series*

The chart below shows the results of the automatic detection of uniquely formatted templates for the first 200 lines of the sample document. The first 65 lines of this automatically generated template number series matches the manually created version shown on Figure 4. Now, we collected enough information for getting to the second subtask: detecting the hierarchical structure in the created number series.

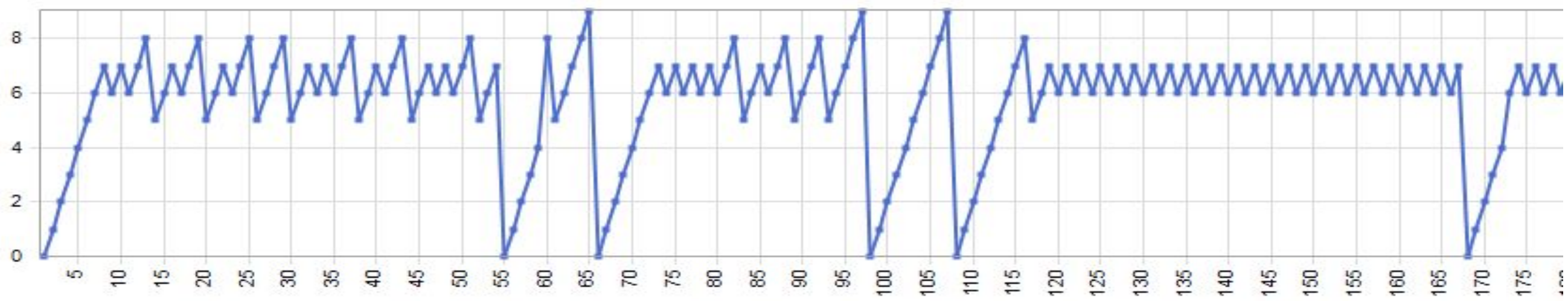

Figure 12. Template number series of the sample financial report presented as a chart.

The fundamental quality that we are looking to detect in this number series derived from our assumption about the correlation between the document formatting and the source data structure and can be defined as "hierarchy of repeating patterns". Although the terms "hierarchy" and "repeating patterns" are well defined, the definition of their combination in our context, is not trivial.

The methodology of identifying hierarchical structure in sequences, introduced by Craig Nevill-Manning and Ian Witten in 1997 [5] offers the SEQUITUR algorithm for building a hierarchical structure from a sequence of discrete symbols by replacing repeated phrases with a grammatical rule that generates the phrase. The SEQUITUR algorithm was designed and successfully used for semantic analysis and packaging of arbitrary text. We will use the same principle of replacing repeated elements, but in the context of a number series and with a different goal of detecting the hierarchical structure of repeating patterns included into each other.

The general definition of the term "repeating pattern" assumes that the same pattern elements could appear in any place of the number series, but in our context, we would be interested only in those that directly follow each other and belong to the *nested hierarchy* (that we are looking to detect). We can define the fundamental quality of this hierarchy as inclusion of repeating patterns. This means that the repeating patterns on each level of the hierarchy (except the lowest one) could be revealed only by removing the multiple instances of the included patterns or replacing them with a single instance.

### *Collapsing repeating patterns*

Let us introduce "collapsing", the action of replacing multiple instances of repeating pattern elements directly following each other with a single instance. The green points on the left part of the image below represent the included repeating patterns, which after collapsing into a single instance reveal the next level of repeating patterns, presented on the right part.

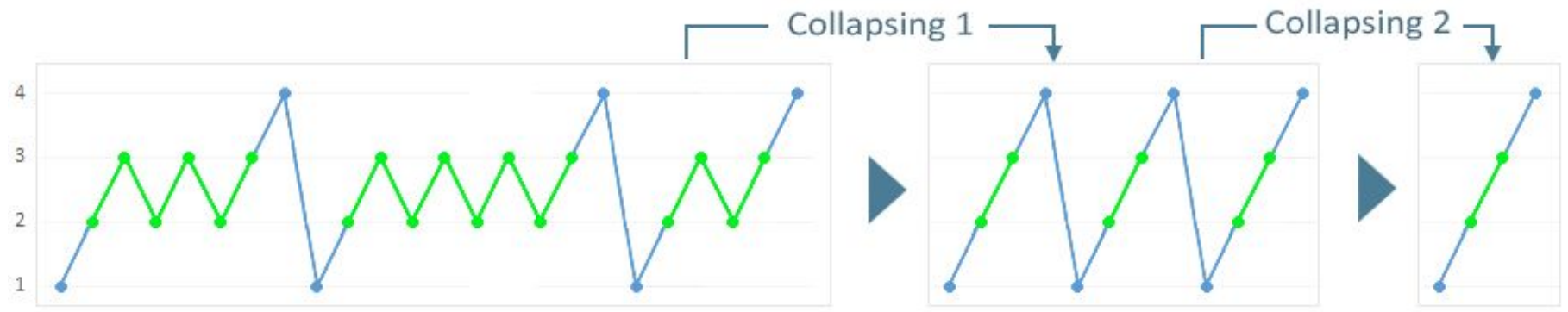

Figure 13. Repeating patterns collapsing procedure.

Applying this procedure to a "pure" hierarchy (single hierarchy without missing elements and no noise), will lead to collapsing the repeating patterns of the inmost (the lowest) level of the hierarchy, thus revealing the repeating patterns of the next level, and so on.

Each repeating pattern could consist of three sequential parts: header elements, included lower level repeating pattern and footer elements. The pattern on each level can be defined by the following recursive transformation:

$$\{p_{i+1} = h_{i+1} + C_i(p_i) + f_{i+1}\}_{\ i = 0 \text{ to } n}$$

where:

| | |
|---|---|
| $i$ | - hierarchical level, |
| $p_i$ | - repeating patterns of the i-th level, |
| $h_i$ | - header of the i-th level, |
| $f_i$ | - footer of the i-th level, |
| $p_0$ | - repeating patterns of the lowest hierarchical level, |
| $p_n$ | - repeating patterns of the highest hierarchical level, |
| $C_i(p_i)$ | - repeating patterns collapsing convolution |

Applying this procedure multiple times will lead to collapsing of the entire structure and converting it into a single instance of the top level repeating pattern.

Below are the definitions summarizing the described process.

1. ***Repeating pattern*** *- identical groups of ascending sequential template numbers directly following each other.*
2. ***Collapsing repeating pattern*** *- replacing multiple instances of repeating patterns with a single one.*
3. ***Child repeating pattern*** *- repeating pattern that after collapsing appears to be a part of another repeating pattern.*
4. ***Parent repeating pattern*** *– a set of templates that turns into a repeating pattern after collapsing the included repeating pattern.*
5. ***Detailed repeating pattern*** *– the collapsible repeating pattern with no child repeating patterns.*
6. ***Hierarchy of repeating patterns*** *– a set of patterns produced by sequential collapsing of included repeating patterns.*

Also, let us introduce a few basic rules limiting the scope of considered documents.

1. *The formatting structure of the document can be described by a superposition of repeating pattern hierarchical structures.*
2. *Parent repeating pattern can have only one child repeating pattern.*

### *Removing Noise*

We implemented the automatic procedure that detects a single repeating pattern hierarchy with the highest number of lines in its detailed (lowest level) section and separates it from the rest of the templates, which are marked as "noise".

This procedure goes through multiple iterations applying the pattern collapsing and "noise" removal routine, starting from the preselected lowest level. During each iteration, it finds the repeating pattern that contains either the starting templates (first iteration only) or the templates of the previously collapsed pattern and applies the collapsing procedure to it, thus reducing the number of number series elements and revealing the repeating pattern of the next level. In the process, it detects the templates that did not match the parent/child hierarchical rules and moves them into a separate group, restarting the iterations after each removal.

The next chart demonstrates the modifications of the templates number series for the pattern-collapsing procedure applied to the templates number series produced by the pattern-recognition step for the first 400 lines of the sample financial report (Figure 14).

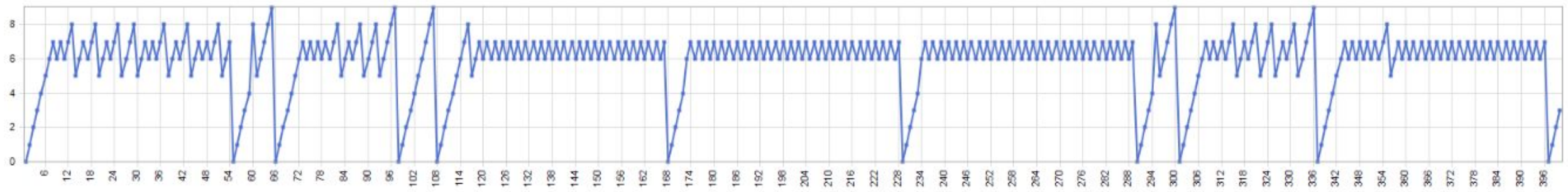

Figure 14. Template number series of the sample financial report for the first 400 lines.

The pattern-collapsing procedure, applied to this number series during the first processing iteration, used the initiating pattern consisting of templates 6 and 7, found during the templates recognition processing as the detailed templates with maximum line count (Table 2).

In the process, the pattern-collapsing procedure goes through the number series identifying the templates that did not comply with the rules of the hierarchical parent-child relations and marking them as "noise" for deletion. Then the procedure deletes all the marked templates and restarts the pattern-collapsing procedure.

The definition of the "noise" templates as "not fitting the hierarchy" assumes that their detection requires revealing of the hierarchical relations between the surrounding templates. At the same time, revealing the parent-child hierarchical relations needs collapsing of the repeating patterns, which, in turn, requires removal of the "noise" templates. Therefore, the collapsing process restarts after each "noise" template's removal.

In our case, the first iteration removed all the "noise" templates, which led to turning the input number series into a pure hierarchy (image below).

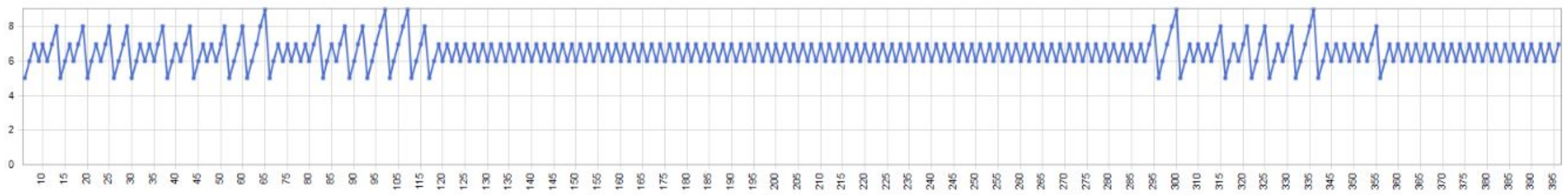

Figure 15. The template number series of the sample financial report with the removed “noise” templates.

The following three charts demonstrate the transformations during the repeating patterns collapsing procedure to the “cleaned up” version of the template number series.

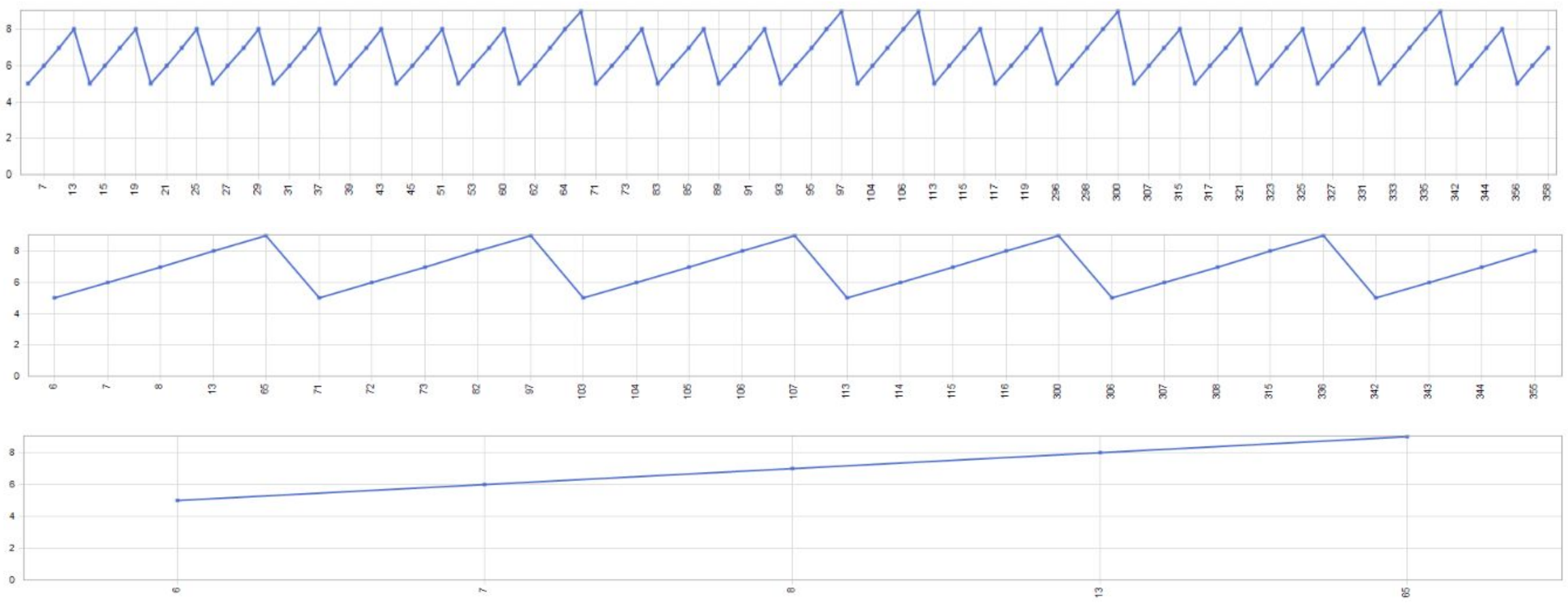

Figure 16. Applying the pattern-collapsing procedure to the “cleaned up” version of the number series.

The first iteration collapsed the lowest level of details (pattern [6, 7], identified by the template recognition step). The next iteration detected and collapsed the pattern [[5, [6, 7], 8], and the last iteration found and collapsed the final repeating pattern [[5, [6, 7], 8], 9].

The results of the automatic procedure matched the manually constructed version and presented as the table (below) or as the two output strings: "[[5, [6, 7], 8], 9]" and "[0, 1, 2, 3, 4]".

| Structure | Level | Pattern | Role |
|---|---|---|---|
| 1 | 0 | [6, 7] | Details level |
| | 1 | [[5, [6, 7], 8] | Summary level 1:  5 - group header, 8 - group footer |
| | 2 | [[5, [6, 7], 8], 9] | Summary level 2:  9 - group footer |
| 2 | 0 | [0, 1, 2, 3, 4] | Page header |

Table 5. The resulting description of the sample financial report’s document formatting structure.

This information, in combination with the rest of the data collected during the described above processing, enables configuring a fully automated data extraction project, which will be the subject of the next chapter.

## Automatic configuration of data extraction projects

The obtained information about the document's structure can be used for automating the data extraction process using some programming environment, but we chose to implement this project by configuring an interactive software tool (SIMX TextConverter 4) via its application programming interface (API). This approach provides a number of advantages including higher manageability and ability to easily employ human expertise for project's monitoring and adjustments at any stage of the process.

Let us start with a brief description of the TextConverter functionality, terminology and the basic configuration workflow. The principle configuration steps include adding and configuring a number of "Templates" responsible for the data extraction (which functionality directly correspond to the same term used in the analytical context).

Each TextConverter template is equipped with a list of fields (input dictionary) configured to perform two functions: (1) - recognizing the specific data segments in the document text flow and (2) - extracting the detected data into the fields of the input dictionary. The input dictionary fields can be directly (visually) or via script connected to the output dictionary fields, thus configuring the data transfer into the connected output database table or a flat file during the project's execution.

The TextConverter templates could be of three types: Detailed, Flat and Hierarchical. The Detailed templates produce output records, while the other two types extract data from the higher levels of the data hierarchy and add it to the generated records.

Each of the non-detailed templates should be configured for extracting data from the header or footer part of the corresponding hierarchical level. The templates should be added and configured in the ascending order of the hierarchical levels, starting from the detailed one (exists by default).

Based on the described workflow, the information necessary for configuring a data extraction process includes:

1. Description of the hierarchical structure of the data.
2. Description of the fields associated with each level of the hierarchy.
3. Description of the roles each group of field would play during the extraction process.

Now, let us use the information obtained during the processing of the sample financial report for configuring the TextConverter data extraction project. The first step would be to add and configure the detailed templates. For that we can take the most inner part of the final pattern produced by the pattern recognition process "[[5, [6, 7], 8], 9]" – templates [6, 7] and the associated fields description to add the two TextConverter Detailed Templates with the names D6 and D7.

After that, we will add all the remaining templates from that line as Flat type, with the role equal to "Header" in case of their positioning to the left of the detailed templates [6, 7], and – "Footer" in case of their positioning to the right.

The final step would be adding the templates from the second pattern line "[0, 1, 2, 3, 4]", excluding those that have a non-body role (header and decor). Below is the screenshot of the fully automatically configured TextConverter project for extracting data from the sample financial report. The project produced the correct output dataset with the processing time on the mediocre PC about 20 seconds.

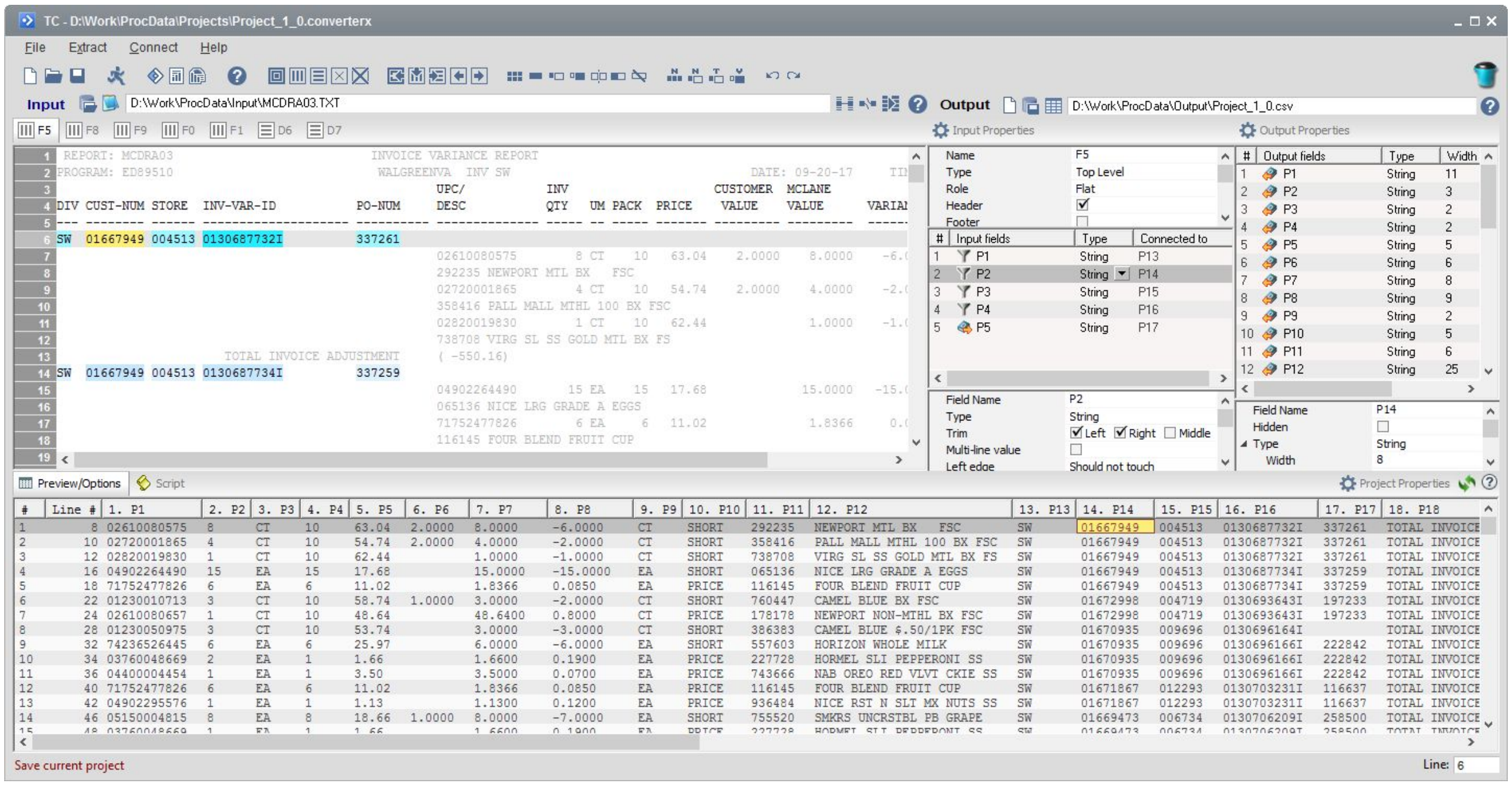

Figure 17. Screenshot of the automatically configured SIMX TextConverter project.

The current version of the described processing engine was implemented using the [SIMX Target Application Development Suite 4.8](SIMX Target Application Development Suite 4.8) and tested against about a hundred documents, most of which were financial reports. Below are several examples of the processing results.

| Format | Document | Template Diagram | Templates | Fields | Structure Strings |
|---|---|---|---|---|---|
| 1 | 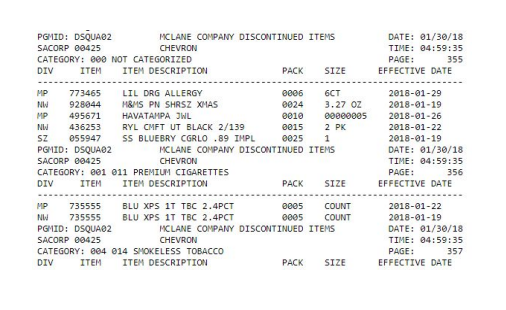 | 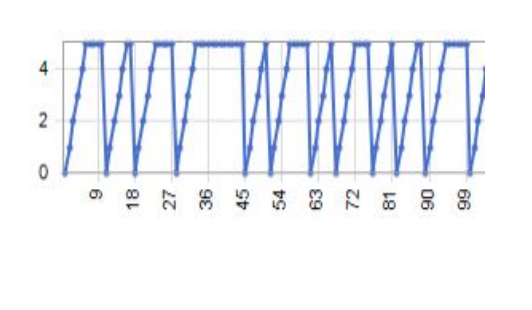 | 6 | 22 | S1:[[0,1,2,3,4],5] |
| 2 | 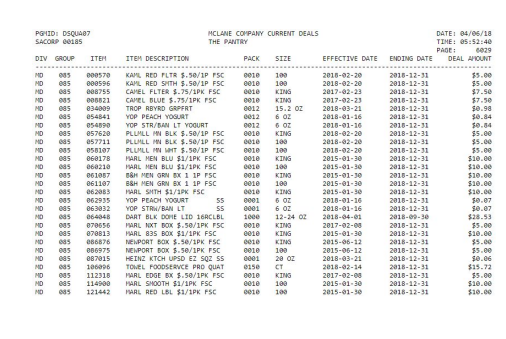 | 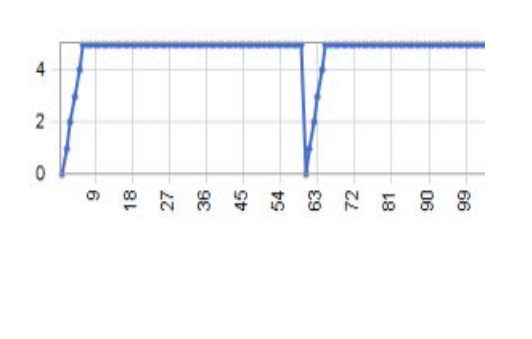 | 6 | 21 | S1: [[0,1,2,3,4],5] |
| 3 | 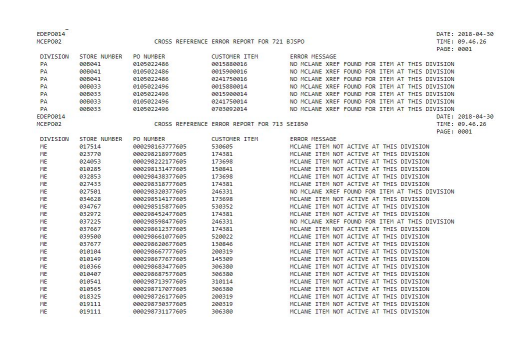 | 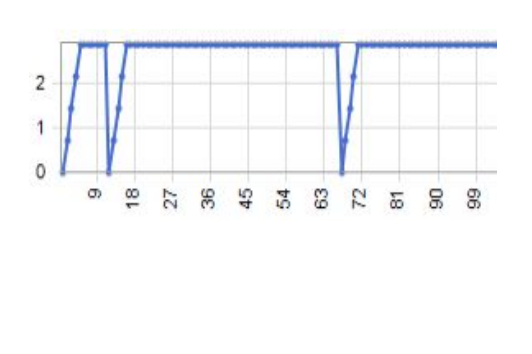 | 5 | 16 | S1: [[0,1,2,3],4] |
| 4 | 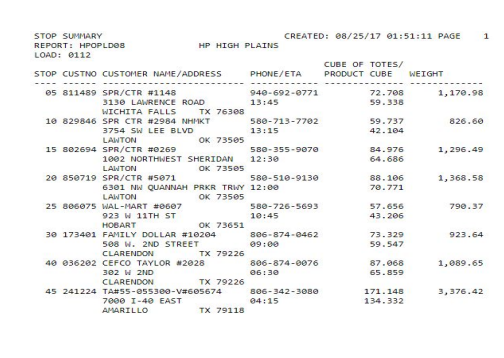 | 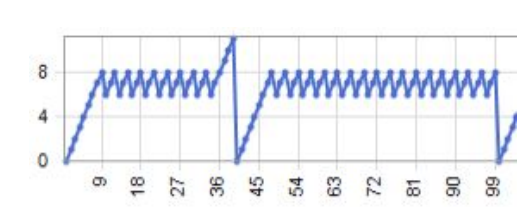 | 11 | 29 | S1: [[0,1,2,3,4,5,[6,7,8],9] |
| 5 | 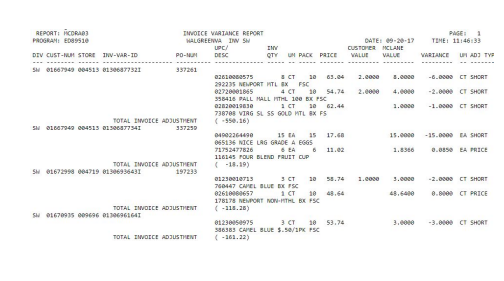 | 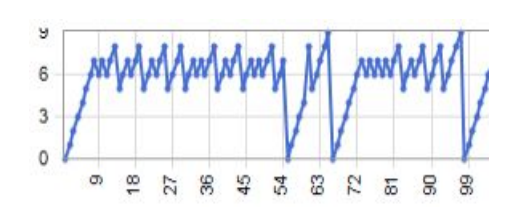 | 10 | 35 | S1: [[5,[6,7],8],9]<br>S2: [0,1,2,3,4] |
| 6 | 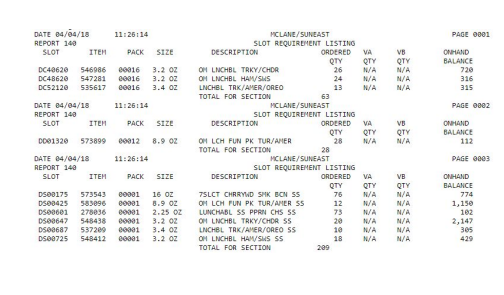 | 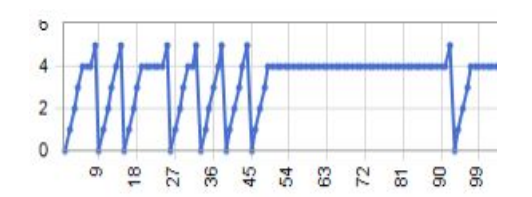 | 6 | 21 | S1: [0,1,2,3,[4],5] |
| 7 | 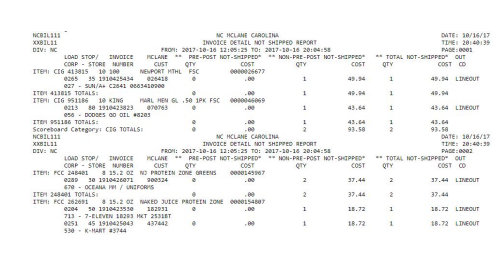 | 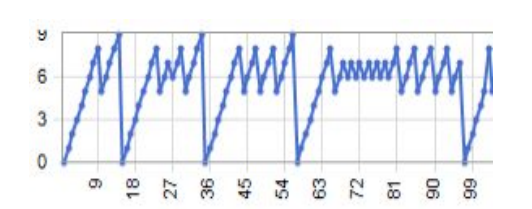 | 10 | 53 | S1: [[5,[6,7],8],9]<br>S2: [0,1,2,3,4] |

Table 6. Examples of the unsupervised data extraction processing.

The automatically configured TextConverter projects provided accurate results for about 95% of the processed files. The produced output tables contained the correct number of records and an accessive number of fields, which included the extra field labels (in addition to the actual data).

# Conclusion

The presented methodology enables unsupervised data extraction from a range of computer-generated documents containing textual data with a single-line formatting. The methodology is based on an automatic detection of repeating patterns in the documents formatting, identifying their hierarchical structure, detecting the data fields and using this information for configuring the fully automated data extraction processing projects.

Although the implemented version of the methodology covers a limited range of documents with a single line formatting, the suggested general approach to the task is applicable for a much wider spectrum of document formatting, including free forms layout with multiline fields.

# Acknowledgment

We would like to express our gratitude to Yuri Yarim-Agaev, Guilherme Pimentel, Boris Althsuler, Veronika Shelyekhova and Mark Davidov for the inspiring discussions and valuable feedback.